\documentclass{iau} 
\usepackage{graphicx}
\usepackage[style=authoryear,backend=bibtex,doi=false,isbn=false,url=false,eprint=false]{biblatex}
\renewbibmacro{in:}{}
\DeclareFieldFormat{pages}{#1}
\DeclareFieldFormat{pages}{\mkfirstpage[{\mkpageprefix[bookpagination]}]{#1}}
\AtEveryBibitem{%
  \clearfield{title}%
  \clearfield{month}%
  \clearfield{eid}%
  \clearfield{number}%
}
\DefineBibliographyStrings{english}{%
  page             = {},
} 

\newcommand{\rosat}{ROSAT}
\newcommand{\xmm}{XMM-Newton}
\newcommand{\chan}{Chandra}

\newcommand{\ero}{eROSITA}

\newcommand{\magoe}{RX J1856.6-3754}
\newcommand{\magzs}{RX J0720.4-3125}
\newcommand{\magos}{RX J1605.3+3249}
\newcommand{\magto}{RX J2143.0+0654}

\newcommand{\jzs}{J0720}
\newcommand{\jos}{J1605}
\newcommand{\jto}{J2143}
\newcommand{\msev}{M7}
\title[\ero\ observations of the M7 INSs]
{Deep \ero\ observations of the \textit{magnificent seven} isolated neutron stars}
\author[Pires, Schwope \& Kurpas]{Adriana Mancini~Pires,
Axel~Schwope \and Jan~Kurpas}
\affiliation{Leibniz Institute for Astrophysics Potsdam (AIP) \\ 
An der Sternwarte 16, 14482, Potsdam, Germany \\ email: {\tt apires@aip.de}}% \\[\affilskip]
\pubyear{2022}
\volume{363}
\jname{Neutron Star Astrophysics at the Crossroads: Magnetars and the Multimessenger Revolution}
\editors{E.~Troja \& M.~Baring, eds.}
\begin{document}
%---------------------------------------------------------------------------------------------------
\maketitle
\begin{abstract}
We report the initial results of deep \ero\ monitoring of the \textit{magnificent seven} isolated neutron stars (INSs). Thanks to a combination of high count statistics and good energy resolution, the \ero\ datasets unveil the increasingly complex energy distribution of these presumably simple thermal emitters. For three targets, we report the detection of multiple (in some cases, phase-dependent) spectral absorption features and deviations from the dominant thermal continuum. Unexpected long-term changes of spectral state and timing behaviour have additionally been observed for two INSs. The results pose challenging theoretical questions on the nature of the variations and absorption features and ultimately impact the modeling of the atmosphere and cooling of highly magnetised neutron stars.

\end{abstract}
%---------------------------------------------------------------------------------------------------
\firstsection 
\section{The \textit{magnificent seven} isolated neutron stars\label{sec_intro}}
\noindent A major outcome of the \rosat\ mission was the discovery of a group of seven nearby radio-quiet isolated neutron stars (INSs; e.g.~\cite{tur09,hab07}). The sources, dubbed the \textit{magnificent seven} (\msev), have been regarded as the closest-to-perfect candidates for testing neutron star emission and cooling models, based on a combination of bright thermal emission, proximity, independent distance estimates, and lack of significant magnetospheric or accretion activity (\cite{pot14}). 
Relative to standard (spin-powered) pulsars, the \msev\ rotate more slowly, show higher thermal X-ray luminosity compared to the spin-down power, and have higher inferred magnetic field intensities. These properties suggest that additional heating of the neutron star crust may have been at work by means of field decay, implying an evolutionary link with magnetars (e.g.~Borghese; these proceedings).
The \msev\ INSs are routinely observed by \ero, the primary X-ray telescope on-board the SRG Observatory (\cite{pre21}).  Moreover, SRG/\ero\ has just successfully completed half of its planned four-year all-sky survey mission: by the time it reaches final depth in late 2023, dedicated identification campaigns for new INS candidates can increase the observed population by significant factors (Kurpas et al.; these proceedings).
%---------------------------------------------------------------------------------------------------
\section{Overview of the \ero\ observations\label{sec_obs}}
%---------------------------------------------------------------------------------------------------
\noindent In the circa two years between first light in October 2019 and the time of this symposium, \ero\ targeted the \msev\ INSs \magzs, \magto\ and \magos\ in eight occasions. In addition, the prototype of the group, \magoe, was observed three times for cross-calibration purposes with the \chan\ LETG HRC-S instrument; these results will be reported elsewhere. In Table~\ref{tab_obs} is an overview of the \ero\ datasets. 
%---------------------------------------------------------------------------------------------------
\begin{table*}[t]
\small
\centering
\begin{tabular}{lccccrrr}
\hline
Target ID & \textit{obsid} & Observation & Net Exposure & GTI & \multicolumn{1}{c}{Counts} & \multicolumn{1}{c}{Rate} & \multicolumn{1}{c}{Remark}\\
RX & & Date & [s] & \multicolumn{1}{c}{[\%]} & \multicolumn{1}{c}{[$\times 10^5$]} & \multicolumn{1}{c}{[s$^{-1}$]} & \\ 
\hline
(1)~\jzs & 700007 &  2019-10-16 & 87810 & 87 & 8.176(15) & 9.4692(17) & XMM TOO\\
(2)~\jto & 700198 &  2019-12-02 & 54650 & 99 & 1.622(3) & 2.979(5) & $p_f=2.46(24)$\,\%\\
     & 720005 &  2020-11-25 & 55390 & 93 & 1.059(3) & 2.822(8) & $p_f=1.9(3)$\,\%\\
     & 730031 &  2021-05-25 & 49400 & 85 & 1.434(3) & 2.903(5) & $p_f=2.72(26)$\,\%\\
     & 740005 &  2021-11-25 & 50400 & 70 & 0.952(3) & 2.806(9) & $p_f=2.1(3)$\,\%\\
(3)~\jos & 720000 &  2020-08-06 & 32360 & 56 & 1.791(3) & 5.585(10) & $\sim$ AO11--AO14\\
     & 730000 &  2021-03-07 & 59460 & 83 & 3.136(3) & 5.380(6) & Spectral change\\
     & 740001 &  2021-09-07 & 54610 & 83 & 1.885(3) & 5.051(9) & Intermediate state\\
\hline
\end{tabular}
\caption{SRG/\ero\ observations of three of the \msev\ INSs.
\label{tab_obs}}
\end{table*}
% --------------------------------------------------------------------------------------------------
All observations were analysed with \textit{eSASS} version 201009, pipeline processing c001, following standard procedures. The net exposures and percentages of good-time-intervals (GTIs) in Table~\ref{tab_obs} are averaged over all active telescope modules; total photons and count rates are in the $0.2-1.5$\,keV energy band. For details on the reduction and spectral analysis of \ero\ data we refer to \cite{sch21}.
%---------------------------------------------------------------------------------------------------
\section{Results and outlook\label{sec_results}}
%---------------------------------------------------------------------------------------------------
\noindent 
The results of phenomenological spectral models are summarised in Table~\ref{tab_spec}; more physically motivated neutron star atmospheres will be investigated elsewhere. We show for each target (IDs as in Table~\ref{tab_obs}) the best-fit column density in units of $10^{20}$\,cm$^{-2}$ ($N_{\rm H,20}$), the temperature $kT_{\rm BB}$ and size of the emission region $R_{\rm BB}$ of the main (eventually, ``cool'') blackbody components, and the energy $\epsilon$ and width $\sigma$ of Gaussian absorption features. The emitting areas are normalised to 300\,pc for \magzs\ and 100\,pc for \magto\ and \magos. For the fit results in (3), $kT_{\rm BB}$ and $R_{\rm BB}$ of the hot blackbody component, as well as the strength of the absorption features (not shown in Table~\ref{tab_spec}), are uncoupled while fitting the spectra of the different epochs together. 
%---------------------------------------------------------------------------------------------------
\subsection{\magzs\label{sec_j0720}}
%---------------------------------------------------------------------------------------------------
Thus far an outlier, the INS is well known to display long-term variability. This behaviour has been interpreted as possibly cyclic and related to the star precession, or caused by an accretion or glitch episode (\cite{hab06,ker07b}).
Coordinated \ero/\xmm\ observations were performed in October 2019. The joint analysis of the observations shows evidence for three absorption features, with energies and properties seemingly consistent with those of previous results reported in the literature (\cite{ham09,bor15}). The dependence of the features with respect to the INS spin is under investigation by means of phase-resolved spectroscopy.
%---------------------------------------------------------------------------------------------------
\subsection{\magto\label{sec_j2143}}
\begin{figure}[t]
\begin{center}
\includegraphics[width=\textwidth]{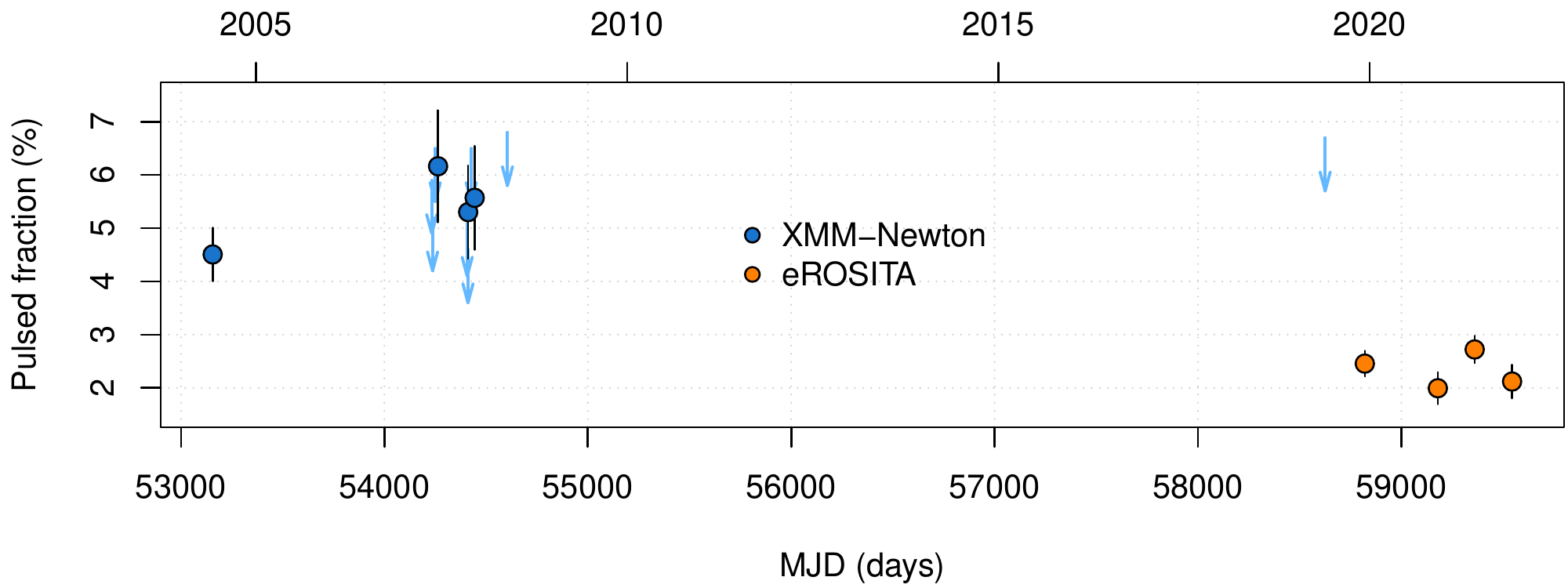}
\end{center}
\caption{\footnotesize Long-term variation of pulsed fraction observed for \magto. \label{fig_pfvar}}
\end{figure}

%---------------------------------------------------------------------------------------------------
The source appears particularly hot in X-rays and bright in the optical/UV (\cite{kap11a}); it may also be the most magnetised in the group if the broad spectral absorption feature at $\sim0.75$\,keV is interpreted as a proton cyclotron line (\cite{zan05}). A recent \xmm\ observation performed in May 2019 failed to detect the smooth flux modulation of the neutron star, previously measured at the 4.5\% level (cf.~detections and upper limits in Fig.~\ref{fig_pfvar}). The decrease in pulsed fraction ($p_f$) is confirmed by four much deeper \ero\ observations, which recover the INS modulation at about the $2.5$\% level (Table~\ref{tab_obs}). 
Moreover, the INS spin is consistently registered in all epochs at a lower frequency than expected from the putative timing solution of \cite{kap09a}: this suggests that the neutron star undergoes a stronger braking than previously thought. 
The analysis of the \ero\ spectra (Fig.~\ref{fig_spec}; \textit{left}) reavealed two additional absorption features and evidence of a colder thermal component (Table~\ref{tab_spec}; see also \cite{schwope09}). Despite the variations in timing behaviour, we measured no significant changes in flux or spectral state over 17\,years. 
%---------------------------------------------------------------------------------------------------
\begin{table*}[t]
\small
\centering
\begin{tabular}{lrrrrrrrrrrr}
\hline
 & \multicolumn{1}{c}{$N_{\rm H,20}$} & \multicolumn{1}{c}{$kT_{\rm BB}^{\rm cool}$} & \multicolumn{1}{c}{$R_{\rm BB}^{\rm cool}$} & \multicolumn{1}{c}{$kT_{\rm BB}$} & \multicolumn{1}{c}{$R_{\rm BB}$} & \multicolumn{1}{c}{$\epsilon_1$} & \multicolumn{1}{c}{$\sigma_1$} & \multicolumn{1}{c}{$\epsilon_2$} & \multicolumn{1}{c}{$\sigma_2$} & \multicolumn{1}{c}{$\epsilon_3$} & \multicolumn{1}{c}{$\sigma_3$}\\
\cline{7-12}
 & \multicolumn{1}{c}{[cm$^{-2}$]} & \multicolumn{1}{c}{[eV]} & \multicolumn{1}{c}{[km]} & \multicolumn{1}{c}{[eV]} & \multicolumn{1}{c}{[km]} & \multicolumn{6}{c}{[eV]} \\
\hline 
(1) & 7.4(5) & & & $85.9_{-0.3}^{+0.4}$ & $5.5(1.0)$ & $359_{-9}^{+7}$ & $63_{-8}^{+9}$ & $548_{-7}^{+4}$ & $<17$ & $770_{-7}^{+8}$ & $35_{-13}^{+14}$ \\
(2) & $13.2_{-0.9}^{+0.8}$ & $43.0_{-2.0}^{+3}$ & $<90$ & $107.4_{-2.8}^{+3}$ & $0.6(3)$ & $392(3)$ & $86(6)$ & $548_{-5}^{+6}$ & $50_{-8}^{+13}$ & $735(6)$ & $32_{-10}^{+13}$ \\
(3) & $5.4(4)$ & $35.9_{-1.3}^{+1.9}$ & $28_{-18}^{+21}$ & $100.4_{-1.2}^{+1.1}$ & $0.96_{-0.29}^{+0.3}$ & $443_{-8}^{+7}$ & $77_{-10}^{+9}$ & $557(6)$ & $36(8)$ & $845_{-12}^{+8}$ & $133_{-13}^{+6}$ \\ 
\hline
\end{tabular}
\caption{\footnotesize Results of the spectral analysis. The observations of Table~\ref{tab_obs} are fitted simultaneously with high-quality EPIC pn exposures of each target. In (3) the uncoupled $kT_{\rm BB}, R_{\rm BB}$ parameters are $105.2_{-3}^{+1.8}$\,eV, $0.74_{-0.26}^{+0.4}$\,km (March 2021), and $102.3_{-2.8}^{+2.4}$\,eV, $0.9_{-0.3}^{+0.4}$\,km (September 2021). 
\label{tab_spec}}
\end{table*}
%---------------------------------------------------------------------------------------------------
%---------------------------------------------------------------------------------------------------
\subsection{\magos\label{sec_j1605}}
\begin{figure}[t]
\begin{center}
\includegraphics[width=0.495\textwidth]{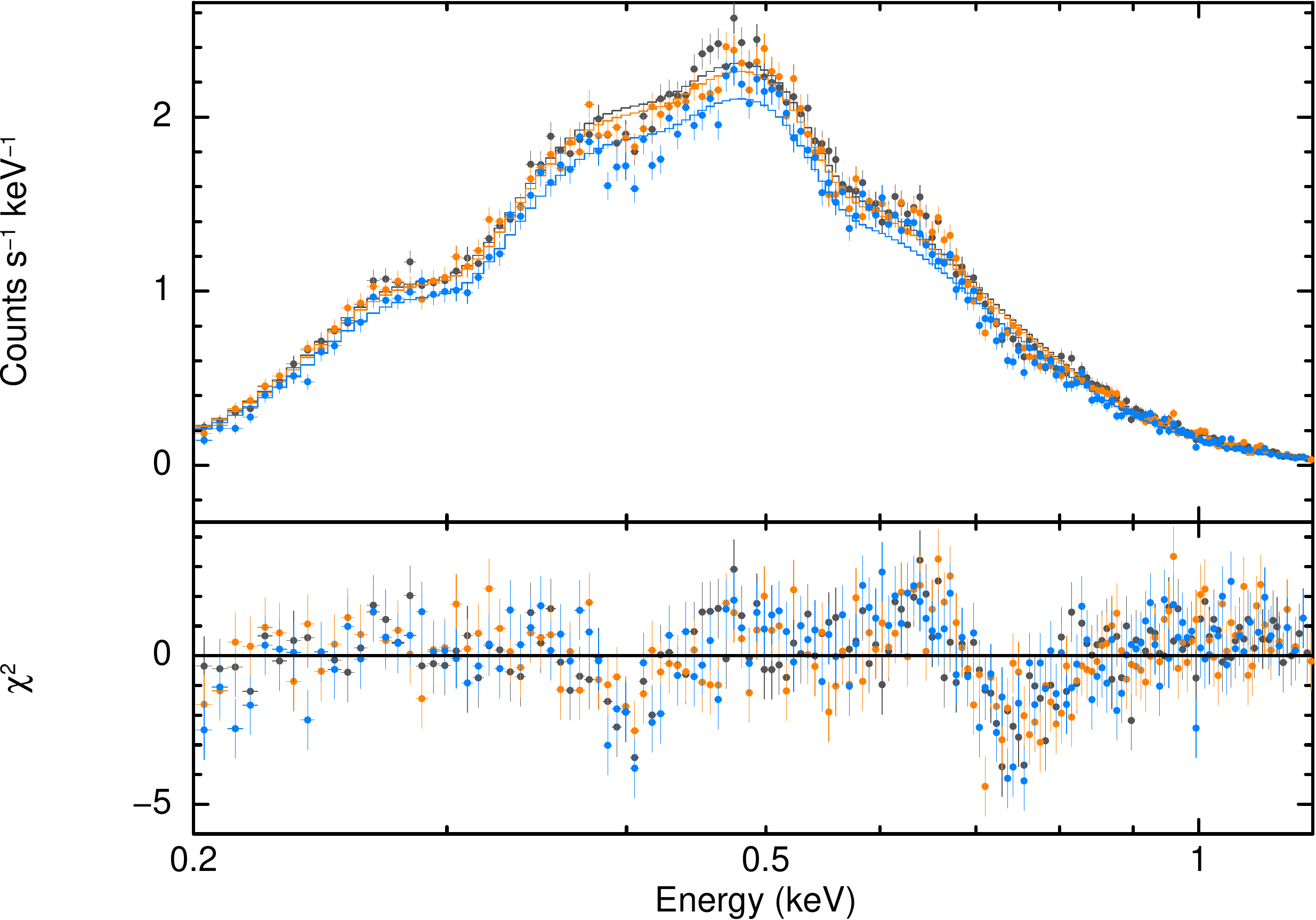}\hspace{0.005cm}
\includegraphics[width=0.495\textwidth]{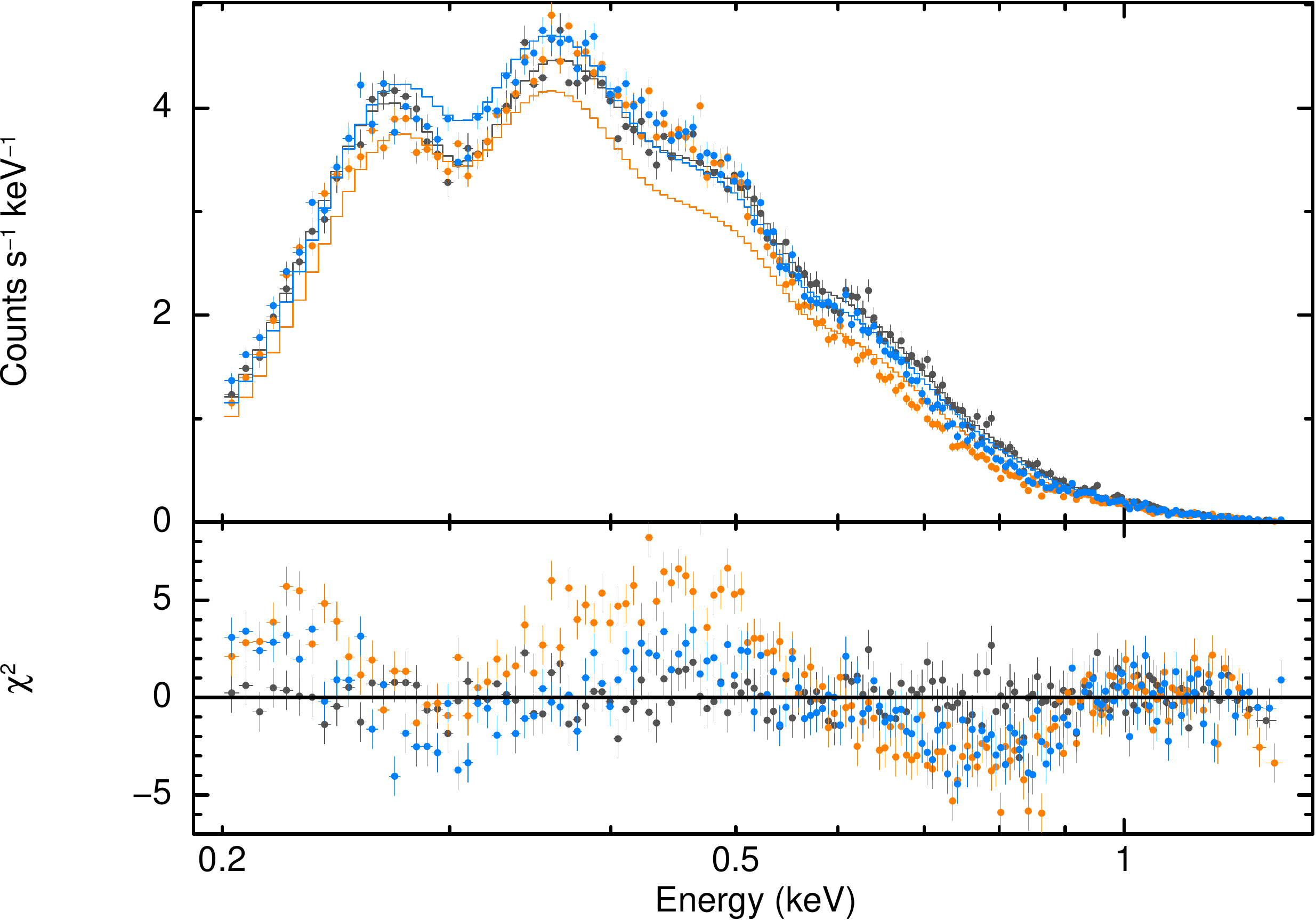}
\end{center}
\caption{\footnotesize \ero\ spectra of \magto\ (\textit{left}) and \magos\ (\textit{right}), where we highlight the spectral changes undergone by the latter and absorption features detected in the spectra of the former. Three epochs are shown for each target (see the text for details). 
\label{fig_spec}}
\end{figure}

%---------------------------------------------------------------------------------------------------
We have recently investigated the INS with \xmm\ (\cite{pir19,pir14}). While the results of a first (August 2020) \ero\ observation are consistent with those of the AO11--AO14 campaigns, in a second observation, conducted seven months later, the source shows very significant variations in its spectral state (cf.~orange as opposed to the dark grey flat residuals in Fig.~\ref{fig_spec}; \textit{right}). The changes can be described by an increase of temperature and decrease of size of the emitting region of the main (hot) blackbody component (see the caption of Table~\ref{tab_spec}), accompanied by an overall $5\%$ decrease in source flux. In addition to the broad feature at 0.4\,keV, other deviations from the thermal continuum are observed, in particular a prominent absorption at 0.8\,keV (cf.~previous reports in \cite{hab07,pir14}). The variations are confirmed by an \xmm\ TOO performed in July 2021 (F.~Haberl, priv.~comm.). A third \ero\ observation from September 2021 indicates that the source is gradually returning to its previous state (blue in Fig.~\ref{fig_spec}; \textit{right}). Interestingly, \cite{pir19} reported a low-significance temporal trend on the parameters of the source in the analysis of RGS data dating back to 2002. 
%---------------------------------------------------------------------------------------------------
%---------------------------------------------------------------------------------------------------
\subsection{Outlook}
%---------------------------------------------------------------------------------------------------
Up to now, the \msev\ have been considered a unique group of cooling neutron stars, sharing a rather stable and well-defined set of properties. The long-term variations observed for now three sources challenge this perception: monitoring of their timing and spectral behaviour will allow the investigation of possible scenarios and the physical mechanisms behind the changes. Forthcoming calibration improvements will likewise help us better determine the column density towards the sources and the properties of the (broad and narrow, spin-dependent) absorption features, which are crucial to tackle their origin. Overall, the first two years of the mission have led to exciting results; we expect \ero's contribution to be paramount not only regarding population properties but also for the modeling of the atmosphere and thermal evolution of highly magnetised neutron stars. 
%---------------------------------------------------------------------------------------------------
\section*{\small References}
\begingroup
\let\clearpage\relax
%
%  These Macros are taken from the AAS TeX macro package version 4.0.
%  Include this file in your LaTeX source only if you are not using
%  the AAS TeX macro package and need to resolve the macro definitions
%  in the BibTeX entries returned by the ADS abstract service.
%
%  If you plan not to use this file to resolve the journal macros
%  rather than the whole AAS TeX macro package, you should save the
%  file as ``aas_macros.sty'' and then include it in your paper by
%  using a construct such as:
%	\documentstyle[11pt,aas_macros]{article}
%
%  For more information on the AASTeX macro package, please see the URL
%	http://www.aas.org/publications/aastex.html
%  For more information about ADS abstract server, please see the URL
%	http://adswww.harvard.edu/ads_abstracts.html
%

% Abbreviations for journals.  The object here is to provide authors
% with convenient shorthands for the most "popular" (often-cited)
% journals; the author can use these markup tags without being concerned
% about the exact form of the journal abbreviation, or its formatting.
% It is up to the keeper of the macros to make sure the macros expand
% to the proper text.  If macro package writers agree to all use the
% same TeX command name, authors only have to remember one thing, and
% the style file will take care of editorial preferences.  This also
% applies when a single journal decides to revamp its abbreviating
% scheme, as happened with the ApJ (Abt 1991).

\def\ref@jnl{}

\def\aj{\ref@jnl{AJ}}                   % Astronomical Journal
\def\araa{\ref@jnl{ARA\&A}}             % Annual Review of Astron and Astrophys
\def\apj{\ref@jnl{ApJ}}                 % Astrophysical Journal
\def\apjl{\ref@jnl{ApJ}}                % Astrophysical Journal, Letters
\def\apjs{\ref@jnl{ApJS}}               % Astrophysical Journal, Supplement
\def\ao{\ref@jnl{Appl.~Opt.}}           % Applied Optics
\def\apss{\ref@jnl{Ap\&SS}}             % Astrophysics and Space Science
\def\aap{\ref@jnl{A\&A}}                % Astronomy and Astrophysics
\def\aapr{\ref@jnl{A\&A~Rev.}}          % Astronomy and Astrophysics Reviews
\def\aaps{\ref@jnl{A\&AS}}              % Astronomy and Astrophysics, Supplement
\def\azh{\ref@jnl{AZh}}                 % Astronomicheskii Zhurnal
\def\baas{\ref@jnl{BAAS}}               % Bulletin of the AAS
\def\jrasc{\ref@jnl{JRASC}}             % Journal of the RAS of Canada
\def\memras{\ref@jnl{MmRAS}}            % Memoirs of the RAS
\def\mnras{\ref@jnl{MNRAS}}             % Monthly Notices of the RAS
\def\pra{\ref@jnl{Phys.~Rev.~A}}        % Physical Review A: General Physics
\def\prb{\ref@jnl{Phys.~Rev.~B}}        % Physical Review B: Solid State
\def\prc{\ref@jnl{Phys.~Rev.~C}}        % Physical Review C
\def\prd{\ref@jnl{Phys.~Rev.~D}}        % Physical Review D
\def\pre{\ref@jnl{Phys.~Rev.~E}}        % Physical Review E
\def\prl{\ref@jnl{Phys.~Rev.~Lett.}}    % Physical Review Letters
\def\pasp{\ref@jnl{PASP}}               % Publications of the ASP
\def\pasj{\ref@jnl{PASJ}}               % Publications of the ASJ
\def\qjras{\ref@jnl{QJRAS}}             % Quarterly Journal of the RAS
\def\skytel{\ref@jnl{S\&T}}             % Sky and Telescope
\def\solphys{\ref@jnl{Sol.~Phys.}}      % Solar Physics
\def\sovast{\ref@jnl{Soviet~Ast.}}      % Soviet Astronomy
\def\ssr{\ref@jnl{Space~Sci.~Rev.}}     % Space Science Reviews
\def\zap{\ref@jnl{ZAp}}                 % Zeitschrift fuer Astrophysik
\def\nat{\ref@jnl{Nature}}              % Nature
\def\iaucirc{\ref@jnl{IAU~Circ.}}       % IAU Cirulars
\def\aplett{\ref@jnl{Astrophys.~Lett.}} % Astrophysics Letters
\def\apspr{\ref@jnl{Astrophys.~Space~Phys.~Res.}}
                % Astrophysics Space Physics Research
\def\bain{\ref@jnl{Bull.~Astron.~Inst.~Netherlands}} 
                % Bulletin Astronomical Institute of the Netherlands
\def\fcp{\ref@jnl{Fund.~Cosmic~Phys.}}  % Fundamental Cosmic Physics
\def\gca{\ref@jnl{Geochim.~Cosmochim.~Acta}}   % Geochimica Cosmochimica Acta
\def\grl{\ref@jnl{Geophys.~Res.~Lett.}} % Geophysics Research Letters
\def\jcp{\ref@jnl{J.~Chem.~Phys.}}      % Journal of Chemical Physics
\def\jgr{\ref@jnl{J.~Geophys.~Res.}}    % Journal of Geophysics Research
\def\jqsrt{\ref@jnl{J.~Quant.~Spec.~Radiat.~Transf.}}
                % Journal of Quantitiative Spectroscopy and Radiative Transfer
\def\memsai{\ref@jnl{Mem.~Soc.~Astron.~Italiana}}
                % Mem. Societa Astronomica Italiana
\def\nphysa{\ref@jnl{Nucl.~Phys.~A}}   % Nuclear Physics A
\def\physrep{\ref@jnl{Phys.~Rep.}}   % Physics Reports
\def\physscr{\ref@jnl{Phys.~Scr}}   % Physica Scripta
\def\planss{\ref@jnl{Planet.~Space~Sci.}}   % Planetary Space Science
\def\procspie{\ref@jnl{Proc.~SPIE}}   % Proceedings of the SPIE

\let\astap=\aap
\let\apjlett=\apjl
\let\apjsupp=\apjs
\let\applopt=\ao

\printbibliography[heading=none]
\endgroup
%---------------------------------------------------------------------------------------------------
\section*{\small Acknowledgements}
\footnotesize
\noindent The authors would like to thank collaborators Konrad Dennerl, Michael Freyberg, Frank Haberl, Anna Karpova, Georg Lamer, Alexander Potekhin, Yuri Shibanov, Valery Suleimanov, Iris Traulsen, and Dmitry Zyuzin for valuable contributions. 
This work is based on data from eROSITA, the soft X-ray instrument aboard SRG, a joint Russian-German science mission supported by the Russian Space Agency (Roskosmos), the Russian Academy of Sciences represented by its Space Research Institute (IKI), the Deutsches Zentrum für Luft- und Raumfahrt (DLR) and the Max-Planck Society, represented by the Max-Planck Institute for Extraterrestrial Physics (MPE).  The eROSITA data shown here were processed using the eSASS software system developed by the German eROSITA consortium.
AMP gratefully acknowledges support from CAS PIFI/2019VMC0008 grant agreement.
This work was supported by the project XMM2ATHENA, which has received funding from the European Union's Horizon 2020 research and innovation programme under grant agreement n$^{\rm o}101004168$.
%---------------------------------------------------------------------------------------------------
\end{document}